\documentclass[12pt]{article}
\usepackage{amsfonts,latexsym,graphicx,amsmath}
\usepackage{stackrel}
\usepackage{doi}
\usepackage{xcolor}
\usepackage{tikz}
\hypersetup{citecolor=red}
\hypersetup{linkcolor=blue}
\hypersetup{urlcolor=blue}

\def\E{\mathcal{E}}

\def\X{\mathcal{X}}
\def\Y{\mathcal{Y}}

\def\B{\mathcal B}
\def\C{\mathcal C}
\def\D{\mathcal D}
\def\U{\mathcal U}
\def\T{\mathcal T}
\def\I{\mathcal I}
\def\H{\mathcal H}

\def\J{\mathcal J}
\def\K{\mathcal K}
\newtheorem{den}{Definition}

\title{Explicit conserved operators for a class of integrable bosonic networks from the classical Yang-Baxter equation}

\author{Phillip S. Isaac, Jon Links, Inna Lukyanenko, Jason L. Werry, \\ 
School of Mathematics and Physics, \\
The University of Queensland, 4072, \\
Australia}

\date{\today}

\begin{document}

\maketitle

\begin{abstract}
Let $B$ denote the weighted adjacency matrix of a balanced, symmetric, bipartite graph. We define a class of bosonic networks given by Hamiltonians whose hopping terms are determined by $B$. We show that each quantum Hamiltonian is Yang-Baxter integrable, admitting a set of mutually commuting operators derived through a solution of the classical Yang-Baxter equation. We discuss some applications and consequences of this result.  
\end{abstract}

\tableofcontents

\newpage
\section{Introduction}

Several studies have seen networks, defined through the use of graphs and hypergraphs, gain prominence in the modelling of quantum phenomena. Examples include the use of {\it star} networks for studies in thermodynamics 
\cite{vvv91}, spin frustration \cite{rv94}, entanglement \cite{hb04}, and information scrambling \cite{l19}. More advanced graph theoretical approaches have been employed in developing protocols for quantum state transfer in spin systems 
\cite{cdel04,kt11}, studying quantum dynamics \cite{adw24}, implementing random graphs in information processing \cite{bfd19}, and for the discretisation of spacetime \cite{g20}.    

The property of {\it integrability} is often encountered in star networks and significant progress has been achieved by exploiting this property.  Arguably, the first example of a quantum system associated with a star network is the Gaudin model for spin interactions \cite{g76}, an example of a {\it central-spin} model. Recent times have seen integrable central-spin models receive renewed attention, particularly with respect to the entanglement properties between the central spin and the surrounding spin bath for the case of $XX$ 
\cite{dmf24,df22,tlpcc23,vcc20,vcppc20,wgl20} and $XY$ \cite{vl23,vl25} interactions.  It is well-understood that the integrability properties in such systems stem from the classical Yang-Baxter equation \cite{bdp05,bd82,j89,k99,s83,s89,s07,s23}. This equation stems from a linearised version of the Yang-Baxter equation, fundamental in the study of integrable quantum systems through the Quantum Inverse Scattering Method 
\cite{f95,r22,s88,tf79}.

The goal of this work is to combine integrability with {\it bosonic} degrees of freedom into the broader picture of networks. This approach is partly motivated by the rising profile of extended Bose-Hubbard models for studies in cold atom systems with dipolar or cavity-mediated interactions
\cite{cblmz25,lmslp09,lps10,lhdlmde16,rbjg24}.
Our investigation below focuses on a class of networks that are associated with bipartite graphs. Examples of this class were introduced in \cite{ytfl17} for cases where the associated bipartite graph $K_{p,q}$ is {\it complete}. (For an introduction to basic notions in graph theory, see \cite{kt11}.) Particular applications have been made for small-sized systems. Adapting the work of \cite{lps10}, physical proposals for the $K_{2,1}$ system through dipolar atoms have been offered in \cite{tywfl20,wytlf18}, and several theoretical aspects of the model have since been studied 
\cite{ccrsh21,cwcrfh24,wcfs22,wcrfh25,wyblf23}.  The $K_{2,2}$ system \cite{tyfl15} has found application in the generation of NOON states \cite{bil23, gwylf22,gywtfl22}, the $K_{3,1}$ system for state transfer on a quantum turntable \cite{ywtfl25}, and both the $K_{3,2}$ and $K_{4,1}$ systems as means for generating $W$-states \cite{bil25}. As shown in \cite{bfil24}, these models on complete bipartite graphs are generally {\it superintegrable}, including the star  models \cite{fl24}; they possess more conserved operators than degrees of freedom. This suggests the possibility to relax the completeness property of the bipartite graph, whilst maintaining integrability.  

Extension to the class of models based on generic bipartite graphs was initiated in 
\cite{l21}.
To define the class of models, let $B$ denote the weighted adjacency matrix of a bipartite graph. We will assume that the graph is {\it balanced}; i.e.,  the vertex sets for the bipartition have equal cardinality. We will also assume that the bipartite graph is {\it symmetric} in the sense of \cite{cm15}; i.e. there exists an involutive graph automorphism that interchanges the vertex sets constituting the bipartition. Under these conditions, $B$ is expressible in the block form 
\begin{align}
B= \left(
\begin{array}{ccc}
0 & | & {\mathcal B} \\
- & & - \\
{\mathcal B} & | & 0 
\end{array}
\right)  
\label{block}
\end{align}
where  ${\mathcal B}$ is a real, symmetric matrix. There are no other constraints placed on  ${\mathcal B}$. 
Let $\{a_j,\,a_j^\dagger:j=1,\dots,m\} \cup 
\{b_j,\, b_j^\dagger:j=1,\dots,m\}$ denote mutually commuting sets of boson operators satisfying
\begin{equation}
\begin{aligned}
&[a_j,\,a_k^\dagger]=[b_j,\,b_k^\dagger]=\delta_{jk}I, \\
&[a_j,\,a_k]=[a_j^\dagger,\,a_k^\dagger]=[b_j,\,b_k]=[b^\dagger_j,\,b_k^\dagger]=0,
\end{aligned}
\label{comms}
\end{equation}
where $I$ denotes the identity operator.
Define the Hamiltonian
\begin{align}
\H&=\U(N^2_a+N^2_b-I)+\sum_{j,k=1}^m 
{\mathcal B}^k_{j}(a_j^\dagger b_k + b_j^\dagger a_k)
\label{ham}
\end{align}
acting on Fock space, where
$ \displaystyle
N_a=\sum_{j=1}^m a_j^\dagger a_j,\, N_b=\sum_{j=1}^m b_j^\dagger b_j$.

A remarkable feature of the model was observed in \cite{l21}.
Since (\ref{ham})
commutes with the number operator 
$N=N_a+N_b$,
it can be block-diagonalised as 
$\displaystyle 
\H = \bigoplus_{{\mathcal N}=0}^{\infty}  \H({\mathcal N})$ where $\H({\mathcal N})$ is the restriction to the Fock subspace on which $N$ acts as ${\mathcal N}I$.
Now, $\H(1)$ acts on a space of dimension $2m$, and is represented as
$$
\H(1)=B 
\cong \left(
\begin{array}{ccc}
{\mathcal B} & | & 0 \\
- & & - \\
0 & | & -{\mathcal B} 
\end{array}
\right) 
$$ 
where the rightmost expression above is with respect to an ordered basis of symmetric and anti-symmetric vectors under the interchange $a_j\leftrightarrow b_j$. This shows that the action 
of ${\mathcal B}$ is embedded into the action of (\ref{ham}) under this basis. It was also claimed in \cite{l21} that (\ref{ham}) is Yang-Baxter integrable. This raises an intriguing question about the relationship between integrability and exact-solvability since ${\mathcal B}$ is an {\it arbitrary} real, symmetric matrix. However, the forms for the conserved operators in \cite{l21} were not made explicit in terms of the coupling parameters that appear in the Hamiltonian. Rather, their existence was deduced through a mapping. The objective of this work is to fill a gap by providing explicit expressions for the conserved operators of (\ref{ham}).

In Sect. 2 we provide preliminary background to formulate the results on the framework of Lie algebras and Poisson structures. Using the classical Yang-Baxter equation allows for the generation of a commutative Poisson subalgebra, representing conservation laws in an abstract dynamical system. This formalism is shown to be realised through the Lie algebra $gl(n)$, and the section concludes with discussion on the problem of quantisation. 
In Sect. 3 these procedures are implemented for a specific solution of the classical Yang-Baxter equation. It is described how this solution is amenable for the construction of (\ref{ham}) on arbitrary bipartite graphs described above.  A route to quantise the commutative Poisson subalgebra is demonstrated. As an application of these results, we establish the integrability of a 4-site Hamiltonian that has been discussed in recent literature \cite{mxjw25}. Final comments are gathered in Sect. 5, while an important technical result on the commutativity of transfer matrices is detailed in the Appendix.

\section{Preliminaries}

\subsection{Lie algebras and related Poisson algebras} \label{pre}
Here we follow the frameworks set out in \cite{al75,t00,v91}.
Let $\mathfrak{g}$ denote a finite-dimensional, complex Lie algebra  with basis $\{X_j:j=1,\dots,d\}$ and structure constants 
$\{c_{jk}^l:j,k,l=1,\dots,d\}$ such that the Lie bracket
\begin{align*}
[X_j,\,X_k]=\sum_{l=1}^d c_{jk}^l X_l
\end{align*}
holds. Let $U(\mathfrak{g})$ denote the universal enveloping algebra of $\mathfrak{g}$ with identity element $I$. 
We remark that $U(\mathfrak{g})$ naturally acquires a Lie algebra structure with commutation relations iteratively evaluated via the Leibniz property; i.e. for $F,G,H\in\, U(\mathfrak{g})$  
\begin{align}
[F,\,GH]=[F,\,G]H+G[F,\,H].
\label{lp}
\end{align}
We say that $F,G\in\, U(\mathfrak{g})$ {\it commute} if 
\begin{align*}
[F,\,G]=0.
\end{align*} 

\begin{den}
Let ${\mathbb C}[\X_1,\dots,\X_d]$ denote the ring of polynomial functions in $d$ indeterminates. Let ${\mathcal A}(\mathfrak{g})$ denote the algebra obtained by defining a Poisson structure on ${\mathbb C}[\X_1,\dots,\X_d]$ through the bracket
\begin{align}
\{f,\,g\}=\sum_{j,k,l=1}^d c_{jk}^l \X_l \frac{\partial f}{\partial \X_j}\frac{\partial g}{\partial \X_k}.
\label{pbr}
\end{align}
\end{den}

\noindent
Note
\begin{align}
\{\X_j,\,\X_k\}=\sum_{l=1}^d c_{jk}^l \X_l
\label{embed}
\end{align}
and that the Leibniz property 
\begin{align}
\{f,\,gh\}=\{f,\,g\}h+g\{f,\,h\}
\label{pleib}
\end{align}
holds. We say that $f,g\in\, {\mathcal A}(\mathfrak{g})$ {\it Poisson-commute} if 
\begin{align}
\{f,\,g\}=0
\label{pcomm}
\end{align}
and a set of elements that pair-wise Poisson-commute is a {\it commutative Poisson subalgebra}.

Let ${\mathcal A}_k(\mathfrak{g})$ denote the space of homogeneous polynomials of degree $k$, satisfying
\begin{align*}
\{ {\mathcal A}_j(\mathfrak{g}),\, {\mathcal A}_k(\mathfrak{g})\}  \subseteq  {\mathcal A}_{j+k-1}(\mathfrak{g})
\end{align*}
as a result of (\ref{pbr}). In particular we have 
\begin{align}
\{ {\mathcal A}_1(\mathfrak{g}),\, {\mathcal A}_k(\mathfrak{g})\}  \subseteq  {\mathcal A}_{k}(\mathfrak{g}).
\label{pad}
\end{align}
Eq. (\ref{embed}) shows that the Lie algebra $\mathfrak{g}$ is embedded in ${\mathcal A}(\mathfrak{g})$ as 
${\mathcal A}_1(\mathfrak{g})$, in that the mapping $\X_j \mapsto X_j$ between basis elements provides a Lie algebra isomorphism. For $\Y_1,\dots,\Y_k\in\, {\mathcal A}_1(\mathfrak{g})$ let the corresponding images under this isomorphism be denoted $Y_1,\dots,Y_k\in\, \mathfrak{g}$. Let $S_k$ denote the symmetric group on $k$ objects. Define the  vector space isomorphism $\iota: {\mathcal A}(\mathfrak{g})\rightarrow U(\mathfrak{g})$ via the following action on products of elements in 
${\mathcal A}_1(\mathfrak{g})$
\begin{equation} 
\begin{aligned}
\iota(\Y_1\dots \Y_k) &=\frac{1}{k!}\sum_{\sigma\in S_k}  Y_{\sigma(1)}\dots Y_{\sigma(k)} ,\\
\iota(1)&= I,
\end{aligned}
\label{viso}
\end{equation}
and extended linearly to all of ${\mathcal A}(\mathfrak{g})$.  Set $U_k(\mathfrak{g})=\iota({\mathcal A}_k(\mathfrak{g}))$. It then follows \cite{v91} 
\begin{align*}
U(\mathfrak{g})=\bigoplus_{k=0}^\infty U_k(\mathfrak{g})
\end{align*} 
provides a $\mathfrak{g}$-module decomposition of the universal enveloping algebra $U(\mathfrak{g})$ in view of (\ref{pad}).

The action 
\begin{align*}
\tilde{\sigma}(a)=-a.
\end{align*}
on $\mathfrak{g}$ is an anti-involution, i.e. $\tilde{\sigma}^2={\rm id}$ and
\begin{align*}
\tilde{\sigma}([a,\,b])=[\tilde{\sigma}(b),\,\tilde{\sigma}(a)], \qquad a,b\in\,\mathfrak{g}.
\end{align*} 
This action extends to $U(\mathfrak{g})$ such that 
\begin{align*}
\tilde{\sigma}(U_k(\mathfrak{g}))=(-1)^k U_k(\mathfrak{g}).
\end{align*}
Now take $F\in U_j(\mathfrak{g})$ and $G\in U_k(\mathfrak{g})$. With respect to commutation relations within $U(\mathfrak{g})$ we have the anti-involution property 
\begin{align*}
\tilde{\sigma}([F,\,G])=[\tilde{\sigma}(G),\,\tilde{\sigma}(F)].
\end{align*}
Define $\sigma=-\tilde{\sigma}$. Then $\sigma$ is an involution, viz. 
\begin{align*}
{\sigma}([F,\,G])=[{\sigma}(F),\,{\sigma}(G)].
\end{align*}
In summary, the action of $\sigma$ given by 
\begin{align}
{\sigma}(U_k(\mathfrak{g}))=(-1)^{k+1} U_k(\mathfrak{g})
\label{sig}
\end{align}
is a automorphism on $U(\mathfrak{g})$ that provides a ${\mathbb Z}_2$-grading with respect to the {\it commutator} action. Note that it is not an automorphism with respect to {\it multiplication} within $U(\mathfrak{g})$, since for example
$\sigma(I)=-I$.

\subsection{Classical Yang-Baxter equation for generating commutative Poisson subalgebras}
Here we follow the framework set out in \cite{h95,s06,s07}. Let $V$ denote a complex vector space of dimension $n$, and let $r(u,v)\in\,{\rm End}(V\otimes V)$ with $u,v\in\,{\mathbb C}$.  Then $r(u,v)$ is termed an $r$-{\it matrix} if it is a solution of the {\it classical Yang-Baxter equation} \cite{bdp05,bd82,j89,k99,s83,s89,s07,s23}
\begin{align}
[r_{12}(u,v),\,r_{23}(v,w)]+[r_{13}(u,w),\,r_{23}(v,w)]
-[r_{21}(v,u),\,r_{13}(u,w)]=0
\label{cybe}
\end{align}
acting on $V\otimes V\otimes V$. Above, the subscripts on $r(u,v)$ denote the embedding of the action within the three-fold tensor product space.  
 
\begin{den}
Given a solution $r(u,v)$ of the classical Yang-Baxter equation (\ref{cybe}), introduce the set of generating elements $\{\T^j_k(u):j,k=1,\dots n\}$,  for an abstract Poisson algebra ${\mathcal P}$. Here, $u$ is a formal variable such that 
\begin{align*}
\T^j_k(u)= \sum_{p=-\infty}^\infty [\T^j_k]_p\, u^p.
\end{align*}
Set
\begin{align*}
\T(u)=\sum_{j,k=1}^n e^k_j \otimes \T^j_k(u).
\end{align*}  
The Poisson bracket is imposed according to  
\begin{align}
\{\T_{1}(u),\,\T_{2}(v)\}
=[\T_2(v),\,r_{12}(u,v)] - [\T_1(u),\,r_{21}(v,u)].
\label{cyba}
\end{align}
In component form this reads
\begin{align*}
 \{\T^j_k(u),\,\T^p_q(v)\}
&=  \sum_{\alpha=1}^n \left( r^{jp}_{k\alpha}(u,v) \T^\alpha_q(v) -  r^{j\alpha}_{kq}(u,v) \T^p_\alpha(v)   
+    r^{p\alpha}_{qk}(v,u) \T^j_\alpha(u)- r^{pj}_{q\alpha}(v,u) \T^\alpha_k(u) \right).
\end{align*}
The Jacobi identity 
\begin{align*}
\{\T^j_k(u),\,\{\T^p_q(v),\,\T^a_b(w)\}\}
+\{\T^p_q(v),\,\{\T^a_b(w),\,\T^j_k(u)\}\}
+\{\T^a_b(w),\,\{\T^j_k(u),\,\T^p_q(v)\}\}=0
\end{align*}
for the Poisson bracket holds as a result of $r(u,v)$ satisfying the classical Yang-Baxter equation (\ref{cybe}). 
\end{den}

Define the elements $ (\T^{(s)})^j_k(u) $ through powers of the generating tensor
\begin{align*}
[\T(u)]^s=\sum_{j,k=1}^n e^k_j \otimes (\T^{(s)})^j_k(u)
\end{align*} 
and set 
\begin{align*}
\I^{(s)}(u)
&=\sum_{j=1}^n(\T^{(s)})^j_j(u).
\end{align*}

Then through use of (\ref{pleib}) and (\ref{cyba}) it is found that for $s>1$
\begin{align*}
\{\T^j_k(u),\,\I^{(s)}(v)\}
&= s \sum_{p,q=1}^n\{\T^j_k(u),\, \T^p_q(v)\} (\T^{(s-1)})^q_p (v) \\
&= s \sum_{\alpha=1}^n\left(r^{jp}_{k\alpha}(u,v) (\T^{(s)})^\alpha_p (v) -   r^{j\alpha}_{kq}(u,v) (\T^{(s)})^q_\alpha (v)\right)  \\ 
&\qquad  + s  \sum_{\alpha=1}^n \left( r^{p\alpha}_{qk}(v,u) \T^j_\alpha(u)(\T^{(s-1)})^q_p (v) -  r^{pj}_{q\alpha}(v,u) \T^\alpha_k(u)(\T^{(s-1)})^q_p (v)\right) 
\end{align*}
leading to 
\begin{align*}
\{\I^{(r)}(u),\,\I^{(s)}(v)\}
&=r \sum_{j,k=1}^n\{\T^j_k(u),\,\I^{(s)}(v)\} (\T^{(r-1)})^k_j (u)\nonumber  \\
&=0. 
\end{align*} 
Formally,  taking the Laurent series
\begin{align*}
 \I^{(s)}(u)= \sum_{j=-\infty}^\infty  \I_j^{(s)}u^j 
\end{align*}
it follows that 
\begin{align}
\{\I_j^{(r)},\,\I_k^{(s)}\}=0, \label{pinvol}
\end{align}
i.e. these elements form a commutative Poisson subalgebra.

\subsection{Shift elements and realisation of the commutative Poisson subalgebra} 
\label{seand}

We take the standard $gl(n)$ basis elements 
$\{E^j_k:j,k=1,\dots,n\}$ with commutation relations  
\begin{align*}
[E^j_k,\,E^p_q]=\delta^p_k E^j_q-\delta^j_q E^p_k
\end{align*}
For later use we introduce the algebra involution
\begin{align}
\theta(E^j_{k})=-E^k_{j},
\label{dual}
\end{align}
extended to all of $U(gl(n))$ such that 
\begin{align*}
{\theta}([a,\,b])=[{\theta}(a),\,{\theta}(b)]
\end{align*}
holds for all $a,\,b\in U(gl(n))$.

Let $\{\E^j_k:j,k=1,\dots,n\}$ denote indeterminates with Poisson bracket
\begin{align}
\{\E^j_k,\,\E^p_q\}=\delta^p_k \E^j_q-\delta^j_q \E^p_k.
\label{pcomms}
\end{align}
Let $B(u)\in {\rm End}(V)$ satisfy
\begin{align}
[B_2(v),\,r_{12}(u,v)]=[B_1(u),\,r_{21}(v,u)],
\label{shift}
\end{align}
which we refer to as a {\it shift} element following \cite{s14}.
It is straightforward to check, as a result of (\ref{cybe},\ref{shift}) that the mapping $\varrho_v: {\mathcal P} \rightarrow {\mathcal A}(gl(n))$ defined by   
\begin{align}
\varrho_v(\T^j_k(u))=B^j_k(u)+\sum_{p,q=1}^n r^{jp}_{kq}(u,v) \E^q_p
\label{preal}\end{align}
preserves the Poisson bracket relation (\ref{cyba}). We refer to $\varrho_v$ as the {\it realisation} of ${\mathcal P}$ afforded by ${\mathcal A}(gl(n))$. This realisation provides the means to generate a commutative Poisson subalgebra in 
${\mathcal A}(gl(n))$ through the images of the elements ${\mathcal I}^{(s)}_k$ under $\varrho_v$.
~~\\

\subsection{Quantisation}

An attempt to quantise the above considerations, by replacing Poisson brackets with commutators, faces a roadblock, as will be explained below. 
\begin{den}
Introduce the set of generating elements $\{T^j_k(u):j,k=1,\dots n\}$, where $u$ denotes a formal variable, for an abstract classical Yang-Baxter algebra $Y$. Set
\begin{align*}
T(u)=\sum_{j,k=1}^n e^k_j \otimes T^j_k(u).
\end{align*}  
The Lie bracket is imposed according to  
\begin{align}
[T_{1}(u),\,T_{2}(v)]
=[T_2(v),\,r_{12}(u,v)] - [T_1(u),\,r_{21}(v,u)].
\label{yba}
\end{align}
In component form this reads
\begin{align}
 [T^j_k(u),\,T^p_q(v)]
&=  \sum_{\alpha=1}^n \left( r^{jp}_{k\alpha}(u,v) T^\alpha_q(v) -  r^{j\alpha}_{kq}(u,v) T^p_\alpha(v)  \right)
\nonumber \\ 
&\qquad + \sum_{\alpha=1}^n \left(   r^{p\alpha}_{qk}(v,u) T^j_\alpha(u)- r^{pj}_{q\alpha}(v,u) T^\alpha_k(u) \right).
\label{ybaa}
\end{align}
\end{den}
The Jacobi identity 
\begin{align}
[T^j_k(u),\,[T^p_q(v),\,T^a_b(w)]]
+[T^p_q(v),\,[T^a_b(w),\,T^j_k(u)]]
+[T^a_b(w),\,[T^j_k(u),\,T^p_q(v)]]=0
\label{ji}
\end{align}
for the Lie bracket holds due to (\ref{cybe}). 

As a result of (\ref{cybe}) and (\ref{yba}), it is seen that the $r$-matrix affords the {\it defining realisation} 
$\pi_v: Y\rightarrow U(gl(n))$ through the homomorphism
\begin{align*}
{\pi}_v(T^j_k(u))= \sum_{p,q=1}^n r^{jp}_{kq}(u,v) E^q_p.
\end{align*}
The shift element equation (\ref{shift}) has the following interpretation in $Y$ \cite{l17}. 
 For $Z(u)\in {\rm End}({\mathbb C}^n)$ define an action on $r(u,v)$ through
\begin{align*}
(Z\circ r)_{12}(u,v)=[Z_2(v),\,r_{12}(u,v)]-[Z_1(u),\,r_{21}(v,u)],
\end{align*}
which has the following skew-symmetry
\begin{align*}
(Z\circ r)_{12}(u,v)=-(Z\circ r)_{21}(v,u).
\end{align*}
Symmetries of $r(u,v)$ are identified as those $Z(u)$ for which the skew-symmetric action is identically zero, 
coinciding with (\ref{shift}).
For any one-dimensional representation $\eta: Y\rightarrow {\mathbb C}$ we have  
\begin{align*}
\eta\left( [T^j_k(u),\,T^p_q(v)] \right) =0.
\end{align*}
It follows that if $B(u)$ satisfies (\ref{shift}), the mapping given by 
\begin{align*}
\eta(T^j_k(u))=B^j_k(u)
\end{align*}
preserves the commutation relations (\ref{ybaa}), thus providing a one-dimensional representation. 
Such one-dimensional representations encode the symmetries of $r(u,v)$ in a generalised sense, and will play a significant role in later calculations. Finally, taking the tensor product of a one-dimensional representation with the defining realisation yields a realisation expressible as  
\begin{align}
\rho_v(T^j_k(u))&=  \eta(T^j_k(u))I + \pi_v(T^j_k(u)) \nonumber \\
&=B^j_k(u)I+ \sum_{p,q=1}^n r^{jp}_{kq}(u,v) E^q_p. \label{real}
\end{align}

Now, following the procedures of Sect. \ref{seand} further by defining
$ (T^{(s)})^j_k(u) $ through powers of the generating tensor
\begin{align*}
[T(u)]^s=\sum_{j,k=1}^n e^k_j \otimes (T^{(s)})^j_k(u)
\end{align*} 
and setting 
\begin{align}
t^{(s)}(u)&=\sum_{j=1}^n(T^{(s)})^j_j(u)
\label{hotm}
\end{align}
it is generally the case that  
\begin{align*}
[t^{(r)}(u),\,t^{(s)}(v)]
&\neq 0.
\end{align*} 
This has been explicitly demonstrated in the study \cite{crt04}. The fundamental issue here is that the operators $T^j_k(u)$ are not commutative, so mimicking the calculations of the previous section that lead to (\ref{pinvol}) fails in general. Methods for circumventing the issues encountered in \cite{crt04} have been investigated in \cite{t06}.

More generally, the problem to quantise  elements of a commutative Poisson subalgebra  to an abelian subalgebra in the universal enveloping algebra of a Lie algebra has been well-studied with a long history, e.g. \cite{ffr94,fm15,mf78,mr19,py08,r06,s07,t00,t02,t03,v91}
A main objective below is to develop techniques suitable for quantising a specific case of commutative, non-homogeneous, polynomial elements in ${\mathcal A}(gl(n))$, where $n$ is even. 

\section{Results}

\subsection{Solution of the classical Yang-Baxter equation, shift element, and the Hamiltonian}
In order to apply the above considerations to investigate the conserved operators for the Hamiltonian (\ref{ham}), we first formulate the appropriate $r$-matrix. 
Let $P$ denote the permutation operator such that
\begin{align*}
P({\mathbf x}\otimes {\mathbf y})={\mathbf y}\otimes {\mathbf x}, \qquad {\mathbf x},\,{\mathbf y}\in {\mathbb C}^n.
\end{align*}
Set 
\begin{align} 
r(u,v)=\frac{u}{2}\left(\frac{1}{u-v}I\otimes I+ \frac{1}{u+v}A\otimes A\right)P,
\label{rm}
\end{align} 
where $A\in {\rm End}({\mathbb C}^n)$.
It may be checked directly that (\ref{cybe}) is satisfied provided 
$A^2=I$.  This solution belongs to a class  discussed in \cite{s06b}.
Next, we turn towards obtaining a non-constant shift element, c.f. \cite{s14}. 
Setting $B(u)=uB$ it is found that  (\ref{shift})
holds provided $AB=-BA$. These conditions are satisfied by choosing $n=2m$ and 
\begin{align}
A&= -I \otimes \sigma^z, \label{a}\\
B&= {\mathcal B} \otimes \sigma^x  \label{b}
\end{align}
for {\it arbitrary} ${\mathcal B}\in\,{\rm End}({\mathbb C}^m)$, where $\sigma^x,\,\sigma^z$ are Pauli matrices. We choose an ordered basis such that $A$ has matrix elements $A^j_k=\delta^j_k (-1)^j$, in which case 
(\ref{rm}) assumes the form
\begin{align*}
r(u,v)&= 
\sum_{j,k=1}^n \frac{u^2(vu^{-1})^{[j+k]}}{u^2-v^2} e^j_k \otimes e^k_j,
\end{align*}
and the relations (\ref{ybaa}) become 
\begin{align*}
 [T^j_k(u),\,T^p_q(v)]
&=    \delta^p_k \frac{u^2(vu^{-1})^{[j+k]}}{u^2-v^2} T^j_q(v) 
- \delta^j_q \frac{u^2(vu^{-1})^{[j+k]}}{u^2-v^2} T^p_k(v)  \\ 
&\qquad  + \delta^p_k\frac{v^2(uv^{-1})^{[p+q]}}{v^2-u^2}T^j_q(u) 
- \delta^j_q\frac{v^2(uv^{-1})^{[p+q]}}{v^2-u^2} T^p_k(u)
\end{align*}
where
\begin{align*}
[z]= \frac{1}{2}(1-(-1)^{z}).
\end{align*}
From (\ref{b}) the matrix $B$ has the property 
\begin{align}
B^{2p}_{2q-1}=B^{2p-1}_{2q}={\mathcal B}^p_q
\label{odd}
\end{align}
and $B^j_k=0$ whenever $j+k$ is even. Note that a re-ordering of the basis brings $B$ into the form (\ref{block}). For later use we make the observation that for even powers of $B$ we have 
\begin{align}
(B^{2l})^{2p-1}_{2q-1}=(B^{2l})^{2p}_{2q}=({\mathcal B}^{2l})^p_q, \qquad l\in\,{\mathbb Z}_{\geq 0}
\label{even}
\end{align}
and $(B^{2l})^j_k=0$ whenever $j+k$ is odd.
  It turns out that the {\it transfer matrix} 
$t(u)\equiv t^{(2)}(u)$ associated with this solution, as defined through (\ref{hotm}), forms a commutative family (see Appendix)
\begin{align}
[t(u),\,t(v)]=0,
\label{ctm}
\end{align}
which is considered the hallmark of Yang-Baxter integrability \cite{f95,r22,s88,tf79}.

As a result of (\ref{real}) we obtain a realisation of the algebra $Y$ through the map
\begin{align*}
\rho_v(T^j_k(u))=uB^j_kI+ \frac{u^2(vu^{-1})^{[j+k]}}{u^2-v^2} E^j_k,
\end{align*}
in turn providing a realisation of the transfer matrix
\begin{align*}
\rho_v(t(u))&=\sum_{j,k=1}^{2m}\rho_v(T^j_k(u)) \rho_v(T^k_j(u)) \\
&=2u^2{\rm tr}(B^2) I+ 
\frac{u^2}{(u^2-v^2)} {\mathfrak C}_1
 +\left(\frac{uv}{u^2-v^2}\right)^2 {\mathfrak C}_2
\end{align*}
where
\begin{align*}
{\mathfrak C}_1&= 2v\sum_{j,k=1}^{2m} B^j_k E^k_j  +\sum_{j,k \,\,{\rm even}}E^j_k E^k_j
+\sum_{p,q\,\,{\rm odd}}E^p_q E^q_p,
\end{align*}
and 
\begin{align*}
{\mathfrak C}_2= \sum_{j,k=1}^{2m} E^j_k E^k_j
\end{align*}
is the second-order Casimir invariant for $gl(2m)$. 

Under the Jordan-Schwinger map
\begin{equation}
\begin{aligned}
E^{2j-1}_{2k-1}&\mapsto a_j^\dagger a_k, & 
\qquad E^{2j}_{2k}&\mapsto a_j^\dagger b_k,   \\
E^{2j}_{2k}&\mapsto b_j^\dagger b_k, & 
\qquad E^{2j}_{2k-1}&\mapsto b_j^\dagger a_k,  
\end{aligned}  
\label{js}
\end{equation}
where the boson operators satisfy (\ref{comms}), the identification
\begin{align*}
2v&\mapsto \U^{-1}  , \\
\frac{1}{2v}{\mathfrak C}_1&\mapsto \H +\U (I +(m-1)N),
\end{align*}
reproduces the Hamiltonian (\ref{ham}).

The above calculations establish that the Hamiltonian (\ref{ham}) is Yang-Baxter integrable, i.e. it is derived from a family of commuting transfer matrices constructed through a solution of the classical Yang-Baxter equation. However, that transfer matrix does not generate any non-trivial conserved operators in addition to the Hamiltonian and total number operator. The situation here is in stark contrast to the familiar setting of Yang-Baxter integrable spin chains \cite{f95,r22,s88,tf79}. There, the degrees of freedom in the system are associated with the lattice sites of the chain. Mathematically, these are incorporated by taking tensor products of vector spaces to build the chain. For (\ref{ham}), tensor products are not part of the construction; the $2m$ degrees of freedom are directly related to the rank of the $gl(2m)$ algebra underlying the construction. In order to obtain more conserved operators we could look for 
``higher-order'' transfer matrices, a topic that has been undertaken for integrable one-dimensional chains \cite{l23,mtv06}. However, these are challenging to construct using the classical Yang-Baxter equation  
\cite{crt04,ffr94}. Instead, the approach taken below is to quantise a set of classical invariant counterparts through the Poisson algebra formulation outlined in the preliminary sections.  

\subsection{Commutative Poisson subalgebra} 

Through (\ref{preal}), we obtain the following realisation for the generators of the Poisson algebra  
\begin{align*}
\varrho_v(\T^j_k(u))=uB^j_k+ \frac{u^2(vu^{-1})^{[j+k]}}{u^2-v^2} \E^j_k
\end{align*}
in terms of the elements satisfying (\ref{pcomms}).
To enable a concise presentation, we introduce the following notational conventions for the calculations hereafter. We use Greek letters to denote odd indices from the set ${\rm o}=\{1,3,\dots,2m-1\}$, and Latin letters for even indices belonging to ${\rm e}=\{2,4,\dots,2m\}$. 
Defining 
\begin{align*}
\E_{\rm e}&=\sum_{j,k\in\,{\rm e}}  e^j_k \otimes \E^k_j +\sum_{\alpha,\beta\in\,{\rm o}}e^\alpha_\beta \otimes \E^\beta_\alpha ,\\
\E_{\rm o}&= \sum_{j\in\,{\rm e}}\sum_{\alpha\in\,{\rm o}} (e^j_\alpha \otimes \E^\alpha_j +e^\alpha_j \otimes \E^j_\alpha)
\end{align*}
allows us to write 
\begin{align}
({\rm id}\otimes \varrho_v)\T(u) &=  uB   + \frac{u^2}{u^2-v^2} \E_{\rm e} + \frac{uv}{u^2-v^2} \E_{\rm o}.
\label{lax}
\end{align} 
We remark that, due to the property (\ref{odd}),  the expression (\ref{lax}) is invariant with respect to the involution 
\begin{align}
\lambda(j)=
\begin{cases}
j-1, & j\in\, {\rm e} ,\\
j+1, & j\in\, {\rm o}.
\end{cases}
\label{invo}
\end{align}

Noting that 
\begin{align*}
\varrho_v(\I^{(s)}(u))&= \sum_{j=1}^n\varrho_v((\T^{(s)})^j_j(u)) \\
&=({\rm tr}\otimes {\rm id})[({\rm id}\otimes \varrho_v)\T(u)]^s  
\end{align*}
it is straightforward to determine the leading-order terms when expanding the above in powers of $u$.
Using the fact that ${\rm tr}(B^s)=0$ when $s$ is odd, which follows from (\ref{block}), the result for $s\geq 2$ is 
\begin{align*}
\varrho_v(\I^{(s)}(u))\sim 
\begin{cases} 
su^{s-1}\,{\rm tr}(B^{s-1}\E_{\rm e}),     &  s \,\,{\rm odd}, \\
\displaystyle  u^s \,{\rm tr}(B^s) + su^{s-2}\left(v\,{\rm tr}(B^{s-1}\E_{\rm o}) 
+\frac{1}{2}\sum_{k=0}^{s-2}   {\rm tr}\left(B^{s-2-k}\E_{\rm e} B^k \E_{\rm e} \right)\right),  &     s\,\,{\rm even}.
\end{cases}
\end{align*}
and in particular 
\begin{equation}
\begin{aligned}
\varrho_v(\I_{s-1}^{(s)}(u))&=
s\,{\rm tr}(B^{s-1}\E_{\rm e}),     &  s \,\,{\rm odd}, \\
\varrho_v(\I_{s-2}^{(s)}(u))&= s\left(v\,{\rm tr}(B^{s-1}\E_{\rm o}) 
+\frac{1}{2}\sum_{k=0}^{s-2}   {\rm tr}\left(B^{s-2-k}\E_{\rm e} B^k \E_{\rm e} \right)\right),  &     s\,\,{\rm even}.
\end{aligned}
\label{expand}
\end{equation}

The images of (\ref{expand}) under (\ref{viso})  provide candidates for constructing the conserved operators for the Hamiltonian (\ref{ham}). Although we only take leading-order terms in the expansion above, we will make remarks in the Discussion to argue why these are sufficient for this purpose.   

\subsection{Conserved operators from the commutative Poisson subalgebra} 
\label{coftcps}

By taking the images of (\ref{expand}) under (\ref{viso}), recalling that $\U^{-1}=2v$, undertaking rescaling, and streamlining notation leaves us to verify that the following operators are mutually commutative:  
\begin{align*}
C(2p)&= \sum_{j,k\in\,{\rm e}}({B}^{2p})^j_{k}E_{j}^k +\sum_{\alpha,\beta\in\,{\rm o}} ({B}^{2p})^\alpha_{\beta}E^\beta_{\alpha} , \\
C(2p+1)&= \widetilde{C}(2p+1)+ \U\sum_{i=0}^{2p} D(2p,i) 
\end{align*}
where
\begin{align*}
\widetilde{C}(2p+1)=\sum_{j\in\,{\rm e}}\sum_{\alpha\in\,{\rm o}}\left( ({B}^{2p+1})^{j}_\alpha E_j^\alpha  + ({B}^{2p+1})_{j}^\alpha E^j_\alpha \right),
\end{align*}
and
\begin{align*}
D(2p,i)=\begin{cases} \displaystyle
\sum_{j,k,r,q\in\,{\rm e}} ({B}^i)_j^{k} ({B}^{2p-i})_r^q E^j_q E^r_k 
+\sum_{\alpha,\beta,\gamma,\delta\in\,{\rm o}} ({B}^i)_\alpha^{\beta} ({B}^{2p-i})_\gamma^\delta E^\alpha_\delta   E^\gamma_\beta  , \qquad i\,\,{\rm even},  \\
\displaystyle
\sum_{j,k\in\,{\rm e}}\sum_{\alpha,\beta\in\,{\rm o}} (({B}^i)_j^\alpha ({B}^{2p-i})_\beta^k+({B}^{2p-i})^\alpha_j ({B}^{i})^k_\beta) 
E^j_k   E^\beta_\alpha , \,\,\,\quad \qquad \qquad i\,\,{\rm odd}.
\end{cases}
\end{align*}
Let $\C(2p),\,\C(2p+1),\,\D(2p,i)\in\,{\mathcal A}(gl(2m))$ denote their Poisson algebra counterparts through $E^a_b\mapsto \E^a_b$ for all $a,b\in\, {\rm e}\cup {\rm o}$.
We have checked, through several pages of long and arduous direct calculations that 
\begin{align*}
[C(2p),\,C(2q)]=[C(2p+1),\,C(2q)]=[C(2p+1),\,C(2q+1)]=0
\end{align*} 
for all $p$ and $q$. Below, we arrive at the same conclusion using symmetry arguments, strongly guided by the approach of 
\cite{v91}.   

A routine calculation verifies that 
\begin{align*}
[C(2p),\,C(2q)]=[\widetilde{C}(2p+1),\,C(2q)]=[\widetilde{C}(2p+1),\,\widetilde{C}(2q+1)]=0.
\end{align*}
We omit the details. 
Now consider 
\begin{align*}
\J=\sum_{i=0}^{2p}[D(2p,i),\,C(2q))].  
\end{align*}
Since
\begin{align*}
C(2q)\in\,gl(2m)  , \qquad \sum_{i=0}^{2p} D(2p,i)\in U_2(gl(2m)) 
\end{align*}
then $\J\in\,  gl(2m)\oplus  U_2(gl(2m)$. 
However, from the Poisson algebra result
\begin{align*}
\sum_{i=0}^{2p}\{\D(2p,i),\,\C(2q))\}=0   
\end{align*}
it follows that the restricted property $\J\in\,  gl(2m)$ must hold.  Applying the automorphism $\sigma$ given by (\ref{sig}) shows that $\sigma(\J)=\J$, while
\begin{align}
\sigma(C(2q))&=C(2q) \nonumber \\ 
\sigma\left(\sum_{i=0}^{2p}D(2p,i)\right)&=
-\sum_{i=0}^{2p}D(2p,i). \label{neg}
\end{align}
This imposes $\J=0$, establishing
\begin{align}
\sum_{i=0}^{2p}[D(2p,i),\,C(2q))]=0.   
\label{cartan}
\end{align}

Next we move to $\K=\K_1+\K_2+\K_3$,  where $\K_j\in\,U_j(gl(2m))$, defined by
\begin{align}
\K &=\sum_{i=0}^{2p}\sum_{j=0}^{2q}[D(2p,i),\,D(2q,j)].
\label{thm}
\end{align}
We may immediately conclude that $\K_3=0$, by consideration of the corresponding Poisson algebra for which
\begin{align*}
\sum_{i=0}^{2p}\sum_{j=0}^{2q}\{{\mathcal D}(2p,i),\,{\mathcal D}(2q,j)\}=0.
\end{align*}
From (\ref{neg}) it follows that $\sigma(\K)=\K$, while on the other hand,
\begin{align*}
\sigma(\K)&=\sigma(\K_1)+\sigma(\K_2) \\
&=\K_1-\K_2. 
\end{align*}
Hence $\K_2=0$. Now, from (\ref{cartan}) we conclude that 
\begin{align}
[\K_1, \,C(2p)]=0
\label{comm1}
\end{align}
as a result of the Jacobi identity (\ref{ji}). Since 
$\displaystyle \sum_{i=0}^{2p}D(2p,i)$ is invariant with respect to the involution (\ref{invo}), so is $\K_1$. Application of the involution (\ref{dual}) shows that, since $\B$ is real and symmetric,
\begin{align*}
\theta\left(\sum_{i=0}^{2p}D(2p,i)\right)=
\sum_{i=0}^{2p}D(2p,i)
\end{align*} 
which in turn yields
\begin{align}
\theta(\K_1)=\K_1.
\label{minx}
\end{align}
Note that there is an embedding 
$\chi: gl(m) \rightarrow gl(2m)$
given by  
\begin{align*}
\chi(E^j_k)=E^{2j-1}_{2k-1}+E^{2j}_{2k}, \qquad\,j,k\in\{1,\dots, m\}. 
\end{align*}
Let ${\rm im}(\chi)\cong gl(m)$ denote the image of $\chi$, which spans the subspace of $gl(2m)$ that is invariant under (\ref{invo}).  Hence $\K_1\in\, {\rm im}(\chi)$.  
It is also clear that $C(2p) \in \, {\rm im}(\chi)$, satisfying 
\begin{align}
\theta(C(2p))&=-C(2p), \label{fun1}\\
[C(2p),\,C(2q)]&=0. \label{fun2}
\end{align}
We say that the symmetric matrix $\B$ is {\it generic} if the eigenvalues of $\B^2$ are distinct.
In such an instance Eq. (\ref{fun2}) indicates that $\{C(2p):p=1,\dots,m\}$ provides a basis for a Cartan subalgebra ${\mathfrak H} \subset {\rm im}(\chi)$, 
Eq. (\ref{comm1}) indicates that $\K_1$ is in the centraliser of ${\mathfrak H}$, and consequently $\K_1\in\,{\mathfrak H}$ since ${\mathfrak H}$ is a {\it maximally} commutative subalgebra. In light of
(\ref{minx}) and (\ref{fun1}) it follows that $\K_1=0$. Thus $\K$ as given by (\ref{thm}) is zero. Consequently 
$[C(j),\,C(k)]=0$ for all $j$ and $k$.
The result also holds for non-generic $\B$, since this can be obtained as the limit of a perturbation. Let  
${\mathbb X}$ denote a matrix that diagonalises $\B$, and ${\mathbb D}$ a diagonal matrix chosen such that
$\B(\epsilon)=\B+\epsilon {\mathbb X}{\mathbb D}{\mathbb X}^{-1}$ is generic in a neighbourhood of $\B$. Since 
$[C(j),\,C(k)]=0$ holds for $\B(\epsilon)$, it also holds in the limit $\epsilon\rightarrow 0$ since the $C(j)$ only depend on the matrix elements of the powers of $\B(\epsilon)$. 

Applying the realisation (\ref{js}) yields the following expressions for the conserved operators of (\ref{ham}):
\begin{align}
C(2p)&= \sum_{j,k=1}^m({\mathcal B}^{2p})^k_{j}(a_j^\dagger a_k + b_j^\dagger b_k ), \label{coneven}\\
C(2p+1)&= \U \sum_{i=0}^{2p} D(2p,i) +\sum_{j,k=1}^m({\mathcal B}^{2p+1})_{j}^k(a_j^\dagger b_k + b_j^\dagger a_k )
\label{conodd}
\end{align}
with
\begin{align*}
D(2p,i)=\begin{cases} \displaystyle
\sum_{j,k,r,q=1}^m ({\mathcal B}^i)^k_{j} ({\mathcal B}^{2p-i})^q_{r}(a^\dagger_j a_q a^\dagger_r a_k +
b^\dagger_j b_q b^\dagger_r b_k), \qquad \,\,\,\qquad i\,\,{\rm even},  \\
\displaystyle
\sum_{j,k,r,q=1}^m \left(({\mathcal B}^i)^k_{j} ({\mathcal B}^{2p-i})^q_{r}+({\mathcal B}^{2p-i})^k_{j} ({\mathcal B}^{i})^q_{r}\right) 
a^\dagger_j a_q b^\dagger_r b_k, \qquad i\,\,{\rm odd}.
\end{cases}
\end{align*}
Note that $N=N_a+N_b=C(0)$ and  $H=C(1)-\U(I+(m-1)C(0))$.

\section{An example}
To illustrate the results, we work through the minimal non-trivial example for which $m=2$. This correspond to the choice $K_{2,2}$ for the bipartite graph. The most general symmetric matrix that can be used to weight the edges of the graph has three independent parameters, as depicted in Fig. 1. For this parametrisation
\begin{align}
\B= -\frac{1}{2}\left(
\begin{array}{cc}\displaystyle
{K}_1 &  \displaystyle  {J}\\
\displaystyle {J} & \displaystyle {K_2}   
\end{array}
\right),  \qquad\qquad J,\,K_1,\,K_2\in\,{\mathbb R},
\label{m2b}
\end{align}
gives the Hamiltonian (\ref{ham}) as 
\begin{align}
\H=\U(N^2_a+N^2_b-I) &- \frac{K_1}{2}(a_1^\dagger b_1 + b_1^\dagger a_1)-\frac{K_2}{2}( a_2^\dagger b_2 + b_2^\dagger a_2) 
- \frac{J}{2} (a_2^\dagger b_1 + b_2^\dagger a_1+ a_1^\dagger b_2 + b_1^\dagger a_2).
\label{4ham}
\end{align}
\begin{figure}[h!] \label{fig:1}
\centering

\begin{tikzpicture}[scale=0.8]

\node[circle, fill=blue, scale=0.7] (l0) at (8, 0) {};
\node[circle, fill=teal, scale=0.7] (l1) at (8, 4) {};
\node[circle, fill=teal, scale=0.7] (l5) at (4, 0) {};
\node[circle, fill=blue, scale=0.7] (l6) at (4, 4) {};
\filldraw[ draw=black, line width=1.5pt]  (l0)--(l1);
\filldraw[ draw=black, line width=1.5pt]  (l0)--(l5);
\filldraw[ draw=black, line width=1.5pt]  (l6)--(l1);
\filldraw[ draw=black, line width=1.5pt]  (l6)--(l5);

\draw (4.35,0.45) node[scale=1.2] {$2$};
\draw (7.65,0.45) node[scale=1.2] {$2$};
\draw (4.35,3.55) node[scale=1.2] {$1$};
\draw (7.65,3.55) node[scale=1.2] {$1$};

\draw (6, 4.8) node[scale=1.2] {$-\displaystyle\frac{K_1}{2}$};
\draw (6, -0.8) node[scale=1.2]  {$-\displaystyle\frac{K_2}{2}$};
\draw (3.3, 2) node[scale=1.2] {$-\displaystyle\frac{J}{2}$};
\draw (8.7, 2) node[scale=1.2]  {$-\displaystyle\frac{J}{2}$};

\end{tikzpicture}

\caption{Graphical presentation of the complete bipartitie graph $K_{2,2}$. The vertex sets coloured blue and teal are assigned labels 1 and 2. Weights are assigned to the edges such that the associated matrix $\B$ given by (\ref{m2b}) is symmetric, leading to the Hamiltonian (\ref{4ham}). }  
\end{figure}

\newpage
Making a  transformation to boson operators $\{c_j,\,c_j^\dagger:j=1,2,3,4\}$ through
\begin{align*}
a_1^\dagger &\mapsto     c_1^\dagger , &  a_1 &\mapsto  c_1, &
a_2^\dagger &\mapsto    c_3^\dagger , & a_2 &\mapsto  c_3, \\
b_1^\dagger &\mapsto    c_4^\dagger , & b_1 & \mapsto  c_4, &
b_2^\dagger &\mapsto    c_2^\dagger , & b_2 &\mapsto  c_2,
\end{align*}
while also implementing
\begin{align*}
\H + \U I-\frac{\U}{2}N^2 &\mapsto H,  &
\U &\mapsto -2U,
\end{align*}
yields 
\begin{align}
H=&-U(c^\dagger_1c_1-c_2^\dagger c_2+c_3^\dagger c_3-c_4^\dagger c_4)^2 
 - \frac{J}{2}(c_1^\dagger c_2 + c_2^\dagger c_1 + c_3^\dagger c_4 + c_4^\dagger c_3  )
\nonumber \\ & 
\qquad  - \frac{K_1}{2}(c_1^\dagger c_4 + c_4^\dagger c_1)  - \frac{K_2}{2}(c_2^\dagger c_3 + c_3^\dagger c_2) .
\label{int}
\end{align}
This Hamiltonian provides an integrable variant of the system discussed in 
\cite{bil23,gwylf22,gywtfl22}. When $K_1K_2=J^2$, i.e. ${\rm det}({\mathcal B})=0$, it coincides with the case considered in \cite{tyfl15}. See also the more recent work \cite{mxjw25}, where integrability of (\ref{int}) was claimed for $K_1=K_2$.


\section{Discussion}

The main objective of this work was to provide explicit forms for the conserved operators of the Hamiltonian 
(\ref{ham}). These conserved operators are expressed through (\ref{coneven}) and (\ref{conodd}). They were obtained by quantisation of classical counterparts obtained through a Poisson algebra approach. 
These results confirm that the models studied in \cite{ytfl17} associated with complete bipartite graphs generalise to models associated with arbitrary bipartite graphs. We remark that this generalised class of bosonic models is distinct from those derived in other studies \cite{s15,s16a,s16b} that are also founded on solutions of the classical Yang-Baxter equation. 

The elements of the commutative Poisson subalgebra were obtained as the leading-order terms in the expansions 
(\ref{expand}). Even though only leading order terms were taken, these are sufficient to provide a set of algebraically independent conserved operators with cardinality equal to $2m$, the number of degrees of freedom for the system. When ${\mathcal B}$ is generic as defined in Sect. \ref{coftcps}, the method of proof follows the lines proposed in \cite{l21}, with detailed analysis to be provided in a future work.   

This finding opens avenues for the investigation of models defined on well-recognised lattice structures. For instance, the case of the square lattice is presented in \cite{l25}.  The explicit form for the Bethe Ansatz solution is described for open, cylindrical, and toroidal boundary conditions. The problem of incorporating an integrable defect into the model is also discussed.

\section{Appendix - Commutativity of the transfer matrices}
To show that (\ref{ctm}) holds, first observe that the left-hand side is a fourth-order polynomial in the $T^j_k(u),\,T^p_q(v)$. 
Using the cyclic rule of trace, Jacobi identity (\ref{ji}), Leibniz property (\ref{lp}), and (\ref{yba}) allows for the order to be reduced step-by-step. How the calculation evolves depends on the order in which the manipulations are taken. Here we sketch out the relevant steps. First
\begin{align*}
[t(u),\,t(v)]&={\rm tr}_{12}([T_1(u) T_1(u),\,T_2(v)T_2(v)]) \\
&={\rm tr}_{12}(T_1(u)T_2(v)[T_1(u),\,T_2(v)] + T_1(u)[T_1(u),\,T_2(v)]T_2(v) \\
&\qquad + T_2(v)[ T_1(u),\,T_2(v)]T_1(u) + [T_1(u),\,T_2(v)]T_2(v)T_1(u)) \\ 
&={\rm tr}_{12}(T_1(u)T_2(v)([T_2(v),\,r_{12}(u,v)] - [T_1(u),\,r_{21}(v,u)] ) ) \\
&\qquad + {\rm tr}_{12}(T_1(u)([T_2(v),\,r_{12}(u,v)] - [T_1(u),\,r_{21}(v,u)] )T_2(v) ) \\ 
&\qquad + {\rm tr}_{12}(T_2(v)([T_2(v),\,r_{12}(u,v)] - [T_1(u),\,r_{21}(v,u)] )T_1(u) ) \\ 
&\qquad + {\rm tr}_{12}(([T_2(v),\,r_{12}(u,v)] - [T_1(u),\,r_{21}(v,u)] )T_2(v)T_1(u) ) 
\end{align*}
which reduces to the compact form
\begin{align*}
[t(u),\,t(v)]
&={\rm tr}_{12}([T_2(v),\,[r_{21}(v,u)T_1(u),\,T_1(u)]]). 
\end{align*}
Using the relations (\ref{yba}) to reduce further to a quadratic expression leads to 
\begin{align*}
[t(u),\,t(v)]={\rm tr}_{12}([[T_1(u),\,r_{21}(v,u)],\,[T_2(v),\,r_{12}(u,v)]]).
\end{align*} 
To advance the calculations to the next stage, we find it useful to incorporate the use of partial transpositions on the second component of the tensor product, denoted $t_2$, leading to  
\begin{align*}
[t(u),\,t(v)]
&= {\rm tr}_{12}( [T_1(u),\, T^{t_2}_2(v)] (r^{t_2}_{21}(v,u) r^{t_2}_{12}(u,v) +r^{t_2}_{12}(u,v)
r^{t_2}_{21})(v,u))  \\
&\qquad -  {\rm tr}_{12}([T_1(u),\, T_2(v)](r_{21}(v,u)r_{12}(u,v) +r_{12}(u,v)r_{21}(v,u)) ). 
\end{align*}
From (\ref{yba}) it follows that 
\begin{align*}
[T_{1}(u),\,T^{t_2}_{2}(v)]
=[r_{12}^{t_2}(u,v),\, T^{t_2}_2(v)] - [T_1(u),\,r^{t_2}_{21}(v,u)]
\end{align*}
which is employed to make the final simplification to
\begin{align*}
[t(u),\,t(v)]=
&= {\rm tr}_{12}(T_2(v) r^{t_1}_{12}(u,v) r^{t_1}_{12}(u,v) r^{t_1}_{21}(v,u)-  T_2(v) r_{12}(u,v)r_{12}(u,v)r_{21}(v,u))   \\
&\qquad +   {\rm tr}_{12}(T_2(v)r_{21}(v,u) r_{12}(u,v)r_{12}(u,v)  -T_2(v) r^{t_1}_{21}(v,u)r^{t_1}_{12}(u,v)r^{t_1}_{12}(u,v))  \\
&\qquad +{\rm tr}_{12}(T_1(u) r_{21}(v,u)r_{21}(v,u)r_{12}(u,v)
- T_1(u) r^{t_2}_{21}(v,u)r^{t_2}_{21}(v,u)r^{t_2}_{12}(u,v)) 
    \\
&\qquad 
+{\rm tr}_{12}(T_1(u)r^{t_2}_{12}(u,v)r^{t_2}_{21}(v,u)r^{t_2}_{21}(v,u)- T_1(u)r_{12}(u,v)r_{21}(v,u)r_{12}(u,v) ) .
\end{align*}
The above vanishes provided 
\begin{align}
[r_{12}(u,v),\,r_{21}(v,u)]&=0, \label{c1}\\
[r^{t_2}_{12}(u,v),\,r^{t_2}_{21}(v,u)]&=0. \label{c2} 
\end{align}
Eq. (\ref{c1}) may be verified from (\ref{rm}) in a direct fashion. Checking the validity of (\ref{c2}) is more involved and relies on
a choice of $A$ satisfying ${\rm tr}(A)=0$, which is the case for (\ref{a}). 

\section*{Acknowledgments}
This research was supported by the Australian Research Council through Discovery Project DP200101339, {\it Quantum control designed from broken integrability}. We thank Angela Foerster for comments and correspondence, and Taras Skrypnyk for helpful feedback. We acknowledge the traditional owners of the land on which The University of Queensland (St. Lucia campus) operates.

\end{document}